# Time-based Fairness Improves Performance in Multi-rate WLANs

Godfrey Tan and John Guttag
*MIT Computer Science and Artificial Intelligence Laboratory*
{godfreyt, guttag}@csail.mit.edu

**Abstract**

The performance seen by individual clients on a wireless local area network (WLAN) is heavily influenced by the manner in which wireless channel capacity is allocated. The popular MAC protocol DCF (Distributed Coordination Function) used in 802.11 networks provides equal long-term transmission opportunities to competing nodes when all nodes experience similar channel conditions. When similar-sized packets are also used, DCF leads to equal achieved throughputs (*throughput-based fairness*) among contending nodes.

Because of varying indoor channel conditions, the 802.11 standard supports multiple data transmission rates to exploit the trade-off between data rate and bit error rate. This leads to considerable *rate diversity*, particularly when the network is congested. Under such conditions, throughput-based fairness can lead to drastically reduced aggregate throughput.

In this paper, we argue the advantages of *time-based fairness*, in which each competing node receives an equal share of the wireless channel occupancy time. We demonstrate that this notion of fairness can lead to significant improvements in aggregate performance while still guaranteeing that no node receives worse channel access than it would in a single-rate WLAN. We also describe our algorithm, TBR (Time-based Regulator), which runs on the AP and works with any MAC protocol to provide time-based fairness by regulating packets. Through experiments, we show that our practical and backward compatible implementation of TBR in conjunction with an existing implementation of DCF achieves time-based fairness.

## 1 Introduction

802.11 is the *de facto* wireless networking standard. In a typical deployment, a mobile node or station equipped with an 802.11 interface communicates over the air to an access point (AP) or base station that is connected to a wired backbone. There are a number of different 802.11 standards. For concreteness, we focus primarily on 802.11b, the most widely used version of 802.11. When multiple mobile nodes wish to use the wireless channel simultaneously, the channel must be apportioned in some "fair" way among them. In 802.11 networks, the apportioning is controlled by DCF at the MAC layer and the queuing mechanism used at the APs.

For reasons we discuss later, nodes connected to 802.11 WLANs transfer data at a number of different rates. So, for example, the channel capacity might have to be apportioned between nodes transferring data at 11 Mbps and nodes transferring data at 1 Mbps. In this paper, we first demonstrate that DCF and the existing queuing schemes at the APs provide a notion of fairness that is inherently inefficient, and then propose and evaluate a better mechanism.

The signal strength and loss rate of indoor wireless channels vary widely, even for nodes that are equidistant from access points [19]. When the 802.11 MAC detects a packet loss (due to the absence of a synchronous *ack*), it continues retransmitting the packet until the maximum retry limit has been reached. However, this is futile when the average signal strength at the receiver is consistently lower than the threshold required for successful packet reception. In such cases, the sender can transmit at a lower data rate (using a more resilient modulation scheme) so that the channel bit error rate (BER) is reduced. In general, there is a trade-off between data rate and BER [11, 16].

Many vendors of APs and client cards implement automatic rate control schemes in which the sending stations adaptively change the data rate based on perceived channel conditions [7, 16, 21]. Many cards also allow users to manually set the data rate. The 802.11b standard defines four different data rates, 1, 2, 5.5 and 11 Mbps respectively. This leads to *rate diversity* in the system, where competing nodes within a cell use different data rates to communicate with the AP (in both uplink and downlink directions). As shown in Figure 1, various data transmission rates were used during 90-minute sessions of a student workshop that took place at MIT. Furthermore, WLANs carry significant amounts of traffic, and thus many APs experience several congested periods. In Section, 3, we discuss the prevalence of rate diversity in more detail.

When multiple nodes are simultaneously exchanging data using different data rates during congested periods, the total network throughput is quite different from what one might expect. Figure 2 illustrates how the aggregate throughput can be dramatically reduced when two competing nodes use different data rates to upload files using TCP. The achieved throughput of the node with the higher transmission rate is reduced by about 3.75 times.

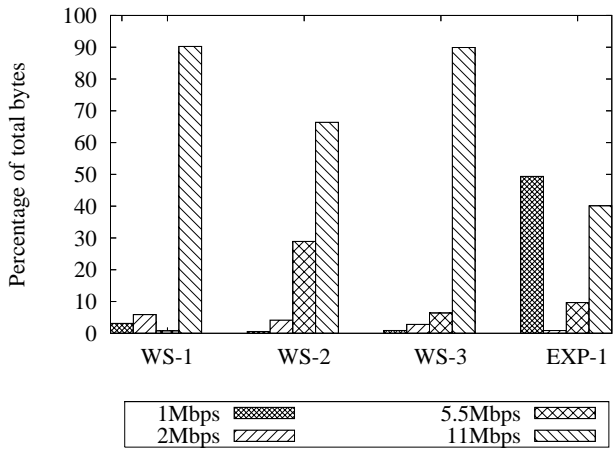

Figure 1: Fractions of bytes transferred at various data rates during three 90-minute workshop sessions (WS) and an experiment (EXP-1).

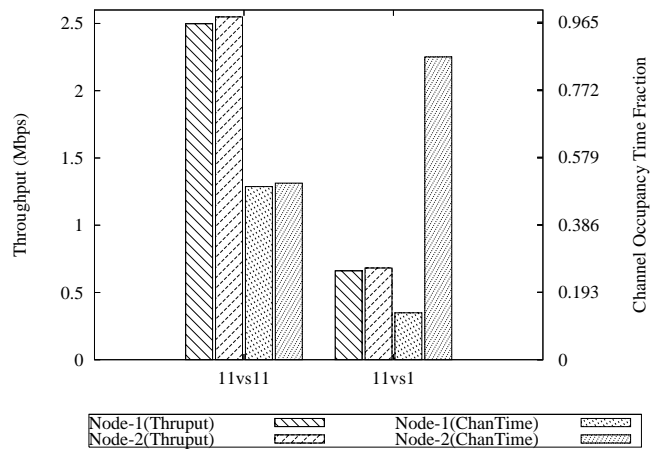

Figure 2: TCP throughputs achieved and fractions of channel occupancy time used by two competing nodes when i) both sending at 11 Mbps and ii) one sending at 11 Mbps and the other at 1 Mbps.

The root cause of this behavior is the definition of fairness used by DCF. This variant of the CSMA medium access protocol is designed to give approximately equal *transmission opportunities* to each competing node. That is to say each node will have approximately the same number of opportunities to send a data frame, *irrespective of the amount of time required to transmit a packet*. When the same-sized packets are used and channel conditions are similar, each competing node, regardless of its data rate, achieves roughly the same throughput, as shown in Figure 2.

Since the node transmitting at 1 Mbps will take several times longer to transmit a frame than the node transmitting at 11 Mbps, the channel is being used most of the time by the slower node. In Figure 2, the fraction of the channel time used by the slower node is $6.4$ times as much as that used by the faster node. Hence, the total throughput is reduced to a level much closer to what one gets when both competing nodes are slow. The faster node pays a penalty for competing against a slow node rather than against another fast node.

Aggregate throughput is also impacted. Naively, one might expect the total throughput of an 11 Mbps and a 1 Mbps channel to be somewhere around $2.93$ Mbps, the average of the total throughputs achieved by a pair of 11 Mbps channels ($5.08$ Mbps) and a pair of 1 Mbps channels ($0.78$ Mbps). However, it is only $1.34$ Mbps, less than half of what one might expect. And the situation is likely to become worse as the emerging 802.11g networks, with a maximum data rate of $54$ Mbps, are deployed alongside relatively slower 802.11b networks. 802.11g users may see far less performance improvement than expected, thus lowering the incentive for users to upgrade to 802.11g cards.

DCF mainly affects the channel capacity allocation in the uplink direction. The packet scheduling mechanism at the AP dictates the channel capacity allocation to clients in the downlink direction. When there are multiple backlogged packets destined to more than one clients, the scheduling scheme must decide the order of transmission. Again, since the channel conditions at the clients vary, different data transmission rates are often used for different clients. Scheduling schemes in the literature [8, 9, 24] provide throughput-based fairness that has been widely-accepted in wired networks and single-rate 802.11 WLAN, in which the data rate for each transmission on the shared medium is the same. When such scheduling schemes are employed at the APs of multi-rate WLANs, the channel capacity allocation on the downlink direction is impacted in a similar undesirable way as in the uplink direction.

We believe that these inefficiencies are best addressed by adopting a notion of fairness that gives each competing client node an approximately equal amount of the shared channel resource: *channel occupancy time*. This notion of of *Time-based* fairness is quite different from the throughput-based fairness notion widely accepted in wired networks and single-rate WLANs. Time-based fairness provides an important property in multi-rate WLANs that throughput-based fairness does not:

**Baseline property:** The long-term throughput of a node competing against any number of nodes running at different speeds is equal to the throughput that the node would achieve in an existing single-rate 802.11 WLAN in which all competing nodes were running at its rate.

I.e., the throughput a node achieves when competing against $n$ nodes is identical to what it would achieve if it were competing against $n$ nodes all using its data rate.

Fairness is, of course, a subjective notion (as any parent of multiple children knows). We do not claim that one notion is "fairer" than the other. However, we do point out that in the presence of rate diversity during congested periods, time-based fairness does improve the overall network performance.

In this paper, we:

- Examine the impact of both time-based and throughput-based fairness on various measures of network efficiency

- Present an analytic framework in which the impact of rate diversity on the network performance is quantitatively evaluated for each fairness notion used

- Validate our model against a deployed 802.11b network

- Show, by collecting and analyzing trace data, that current 802.11b networks indeed suffer the predicted performance degradation in the presence of rate diversity

- Present an effective and efficient scheme, TBR (for Time-based Regulator), for deploying time-based fairness in existing AP-based WLANs, irrespective of the MAC protocol used

- Describe an efficient 802.11-based implementation of TBR that requires changing only the driver on the access point, and

- Demonstrate the relative advantage of time-based fairness, both analytically (using our model) and experimentally (using the 802.11-based implementation)

The rest of this paper is organized as follows. Section 2 analyzes the expected performance impact of both notions of fairness and examines which notion of fairness DCF achieves under various circumstances. Section 3 presents network trace analyses and experiments that demonstrate that rate diversity is common in today's networks. Section 4 describes in detail our scheme to achieve the time-based fairness, Section 5 evaluates our scheme's performance and Section 6 discusses related work.

## 2 Analysis

In this section, we argue why time-based fairness is desirable in some cases and analyze the achieved throughputs of competing nodes, possibly using different data rates and packet sizes, in 802.11-like CSMA WLANs.

### 2.1 Impact of Fairness Notions on Efficiency

We now examine how different notions of fairness impact the overall efficiency of multi-rate WLANs. The measure of fairness between nodes $i$ and $j$ with equal priorities during an interval $(t_1, t_2)$ is: $|\alpha_i(t_1, t_2) - \alpha_j(t_1, t_2)|$, where $\alpha_i(t_1, t_2)$ and $\alpha_j(t_1, t_2)$ are their achieved portions of the shared resource. In this paper, we only focus on nodes with equal priorities. Different notions of fairness are captured by differing definitions of $\alpha$. Let $\alpha_i^t(t_1, t_2)$ and $\alpha_i^r(t_1, t_2)$ be the channel occupancy time and the achieved throughput respectively of node $i$ during $(t_1, t_2)$.

The choice of fairness notion dictates how the network allocates the shared resource (in our case channel capacity) during periods in which demand exceeds supply. The overall network performance as well as the performance of individual nodes can be greatly affected by it. We define network efficiency as the sum of the utility of each competing node based on their shares of shared resource. We use two traffic models, a *fluid model* [8, 27] and a *task model* [4], to examine the impact of fairness notions on overall network efficiency.

In the fluid model, there is a finite number of flows, each of which continuously transfers infinite streams of bits. The network efficiency can be evaluated using its (average) aggregate sustained throughput (*AggrThruput*). Note that while the instantaneous throughput of a node will vary depending upon the its data rate, the expected instantaneous total throughput is time invariant.

In the task model, there is a finite number of flows, each of which transfers a finite number of bits. Since we are providing fairness only among competing nodes, we assume that each node has one flow. In this model, the instantaneous aggregate throughput varies with the remaining task mix. Thus, it is more appropriate to look at network efficiency in other ways such as the average task completion time, *AvgTaskTime*, and the final task completion time, *FinalTaskTime*. Short *AvgTaskTime* is especially desirable for mobile nodes since those that have completed their communication tasks can turn-off their wireless cards to save energy or move to another place to go on with their work. Short *FinalTaskTime* is also desirable since it implies higher long term average aggregate throughput and thus the network can potentially accommodate more tasks.

Figures 3(a) and 3(b) compare the achieved TCP throughputs and the channel occupancy times of two competing nodes when different fairness notions (RF and TF) are used. These figures assume a flow model or a task model in which no flow has yet completed. The graphs are based on the experiments we conducted. In the remainder of this section we demonstrate that these experimental results are consistent with analytical predictions.

Observe that when both nodes transmit at the same rate (11vs11 and 1vs1), the allocations of both throughputs and channel occupancy times are identical for both fairness notions. However, when one node ($n1$) transmits at 1 Mbps and the other ($n2$) at 11 Mbps (see middle bars in figures), nodes achieve equal throughputs under throughput-based fairness, but $n2$ achieves more throughput than $n1$ under time-based fairness. The situation is reversed with respect to the allocation of channel occupancy time. Each node achieves an equal amount of channel occupancy time under time-based fairness, but $n1$ gets a much larger share than $n2$ under throughput-based fairness.

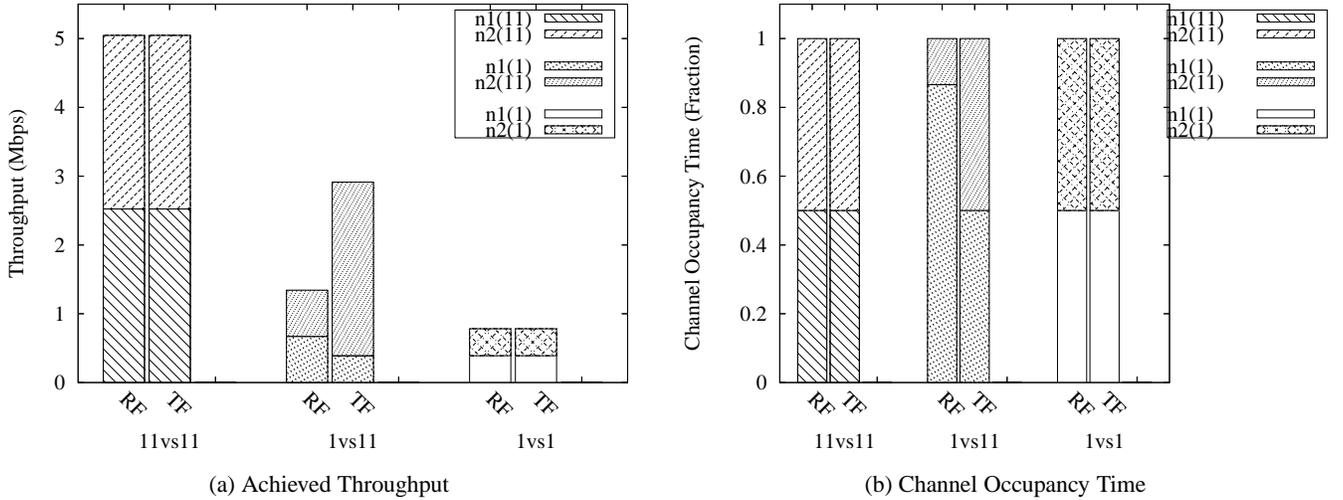

(a) Achieved Throughput  (b) Channel Occupancy Time

Figure 3: Achieved TCP throughputs and fractions of channel occupancy time of two competing nodes in three different combinations of data rates: 11vs11, 1vs11 and 1vs1. *throughput-based fairness* and *TF* denote the throughput-based and time-based fairness notions respectively. E.g. in 3(a), *n1(11)* denotes the throughput achieved by node *n1* transmitting at 11 Mbps.

| Criteria | Measure | RF | TF |
|---|---|---|---|
| Fairness | $\lvert \alpha_i^r(t_1,t_2) - \alpha_j^r(t_1,t_2)\rvert$ | Better | Worse |
|  | $\lvert \alpha_i^t(t_1,t_2) - \alpha_j^t(t_1,t_2)\rvert$ | Worse | Better |
| Efficiency (task model) | *FinalTaskTime* | Same Worse | Same Better |
|  | *AvgTaskTime* |  |  |
| Efficiency (fluid model) | *AggrThruput* | Worse | Better |

Table 1: Comparison of different measures when the throughput-based (RF) and time-based (TF) fairness notions are enforced.

Compared to throughput-based fairness, time-based fairness benefits faster nodes at the expense of slower nodes. However, the fairness property captured by the *baseline property* of Section 1 is maintained. Each class of node performs as it would in a single-rate 802.11 WLAN. For instance, the achieved throughput of $n1$ in both 1vs11 and 1vs1 cases is the same under time-based fairness. The same statement can be made for other performance measures such as per-packet latency.

Table 1 compares various measures of fairness and efficiency for scenarios in which nodes within a cell compete using different data rates. As explained in the rest of this section, the conclusions captured in this table hold for any number of nodes. However, for concreteness, we use the 1vs11 case as a concrete example. Under the task model, we assume that each node has an equal amount of data to transfer. Technically, the same results apply so long as each node has a similar distribution of task size.

When the fluid traffic model is used, higher *AggrThruput* results under time-based fairness as evident in Figure 3(a). *FinalTaskTime* remains unchanged under the task model since the network is work-conserving under both fairness notions. However, *AvgTaskTime* under time-based fairness is lower than that under throughput-based fairness. Under throughput-based fairness, *AvgTaskTime* = *FinalTaskTime*, since both tasks complete at the same time. Under time-based fairness, in contrast, *AvgTaskTime* < *FinalTaskTime*. This is because the task of the 11 Mbps node will complete sooner, since it achieves higher throughput while the completion time of the 1 Mbps node remains the same.

The rest of this section examines how well existing 802.11's DCF achieves each notion of fairness, and presents an analytical framework to predict the network performance in multirate WLANs.

## 2.2 Fairness in AP-based WLANs

Traditional fair queuing algorithms designed for wired networks attempt to provide a fair allocation of the bandwidth on a shared link [8, 9, 24]. Previous work on Fair scheduling in wireless networks generally adopted this notion of fairness [20, 22, 27]. However, unlike wired links, typical wireless networks are half-duplexed in that the channel needs to be shared for both transmitting and receiving packets.

In AP-based WLANs, each AP is just a facilitator and thus the resource used by it to transmit packets destined to a client should be accounted as part of the resource used by the client or its flow. In the rest of this paper, we focus on providing fair channel time shares among competing nodes, not flows. The channel time used by a competing node is the total chan-nel time used in both transmitting and receiving packets to and from the AP. We believe that this notion is more intuitive than the traditional notion of providing fair resource allocations among competing flows. The latter is more suitable for wired networks and *ad hoc* wireless networks, where there are no facilitators present (e.g. when the medium is shared by nodes in a distributed manner) or the facilitator is the only one transmitting on the medium (e.g. router scheduling packets to transmit on an output link).

## 2.3 Network Model

In this subsection, we describe the network model that we use to analyze the performance of AP-based WLANs. In these infrastructure-based WLANs, each wireless node only communicates directly with an AP in order to exchange data with another node inside or outside of the WLAN.

As in much of the existing literature [8, 27], we base our analysis on the fluid traffic model, and thus are concerned with *AgrThruput*. However, the results in this section clearly indicate that when the task traffic model is used, the network efficiency in terms of *AvgTaskTime* is better under time-based fairness than under throughput-based fairness (see Section 2.1).

Let $I$ be the set of competing nodes and $n$ its cardinality. We define $d_i$ and $s_i$ as the data rate used and data packet size used by node $i$. For simplicity of analysis, we assume that $d_i$ and $s_i$ apply to data packets in both uplink and downlink directions of node $i$.

We define the *channel occupancy time* $T(i)$, $0 \leq T(i) \leq 1$, of node $i$ as the fraction of time a wireless node $i$ is able to access the channel to either transmit or receive packets to and from the AP. The channel occupancy time necessary to transfer a data packet includes i) the transmission time of the data packet, ii) the transmission time of a synchronous *ack*, iii) the propagation delays, iv) the inter-frame idle periods necessary for a node to be idle before accessing the channel, and v) the amount of time required to perform retransmissions when necessary. Since we assume that the channel is busy all the time:

$$\sum_{i \in I} T(i) = 1 \qquad (1)$$

Let $R(I)$ and $R(i)$ be the total throughput achieved by all nodes in $I$ and the achieved throughput of node $i$ respectively. We can express $R(i)$ in terms of $T(i)$ as:

$$R(i) = T(i) \times \gamma(d_i, s_i, I) \qquad (2)$$

where $\gamma(d_i, s_i, I)$ is the *baseline throughput* (that nodes experience) as a function of $d_i$, the data rate, and $s_i$, the packet size, holding all else equal. The baseline throughput $\gamma(d_i, s_i, I)$ equals the maximum total achieved throughput when all nodes ($I$) use the same packet size and data rate under similar loss characteristics. For instance, when two nodes simultaneously transfer files using 1500-byte TCP packets and a data rate of 11 Mbps, the baseline throughput (as shown in Figure 2) is 5.08 Mbps. However, the actual throughput $R(i)$ node $i$ depends upon the fraction of time $i$ was able to access the channel, $T(i)$. The total actual throughput of the network is simply:

$$R(I) = \sum_{i \in I} R(i) \qquad (3)$$

Baseline throughput increases with the increase in data transmission rate as well as packet size. The latter is due to reduced per-packet overhead as a result of the larger number of payload bits per packet. By expressing $R(i)$ in terms of $\gamma(d_i, s_i, I)$, we avoid dealing directly with other factors that affect the throughput such as the back-off periods and physical layer overhead, that are independent of the work covered in this paper. $\gamma(d, s, I)$ can be obtained both theoretically and experimentally. In Section 2.7, we report measured values of $\gamma(d, s, I)$ for various values of $d$. Furthermore, we do not deal with varying loss characteristics since our goal is in understanding how diverse data rates and packet sizes affect the network performance.

## 2.4 Impact of DCF on Fairness Notions

802.11's DCF (Distributed Coordinating Function) is far-and-away the most commonly used contention resolution method in 802.11 networks. Although an alternative Point Coordinating Function (PCF) exists, it is not implemented by most AP vendors because of its complexity and issues of co-existence with DCF-based networks. DCF gives equal transmission opportunities (or long-term channel access probability) to each contender [17, 26].

Therefore, competing nodes attempting to send data packets to the AP over the same time interval will be able to transmit equal numbers of frames. DCF's transmission opportunity based mechanism provides fair allocations of both throughput and channel occupancy time only if all contending nodes i) use the same date rate, ii) use the same packet size, and iii) experience very similar loss characteristics. If only the last two conditions hold, DCF achieves throughput-based fairness but does not achieve time-based fairness. For any other combination, DCF achieves neither time-based fairness nor throughput-based fairness.

Figure 4 shows the throughputs achieved by three competing nodes that are either sending or receiving data using the maximum data rate of 11 Mbps and the maximum packet size of 1500 bytes. In uplink directions, the throughput achieved by each node is approximately equal due to DCF. In downlink directions, the throughput achieved by each node is approximately equal largely due to the AP queuing scheme, which usually transmits to wireless clients in a round-robin manner. TCP throughputs are significantly less than UDP throughputs because the transmission overhead of TCP *ack* packets. The total throughputs achieved in the uplink direction are higher than those in the downlink direction. This is because one 802.11 sending node (the AP) cannot fully utilize or saturate the channel since a transmitting node is required to back-off for a random period, between 0 and 610 us, after every successful packet transmission. This overhead is reduced with the increase in number of competing nodes.

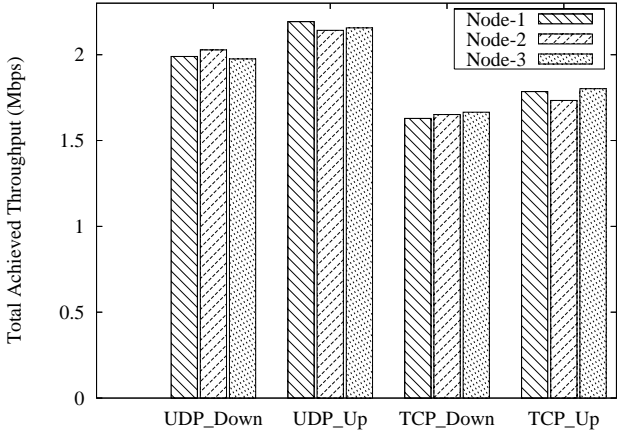

Figure 4: UDP and TCP throughputs achieved by three competing nodes (Cisco-350 cards) each of which is exchanging data at 11 Mbps with a common AP (Cabletron Roamabout-2000). "Up" and "Down" x-axis labels denote that the nodes are sending data to and receiving from the AP respectively.

We now derive the general expression of $T(i)$, the fraction of time node $i$ is able to transmit or receive packets under DCF. For ease of notation, we will use $\gamma_i$ in place of $\gamma(d_i, s_i, I)$. For steady state performance, we can assume that in each round, each competing node transfers a single packet. Thus, $T(i)$ is simply the ratio of the time required for node $i$ to transfer a data frame, which is $\frac{s_i}{\gamma_i}$, to the total time required for every node in $I$ to transfer a data frame.

$$T(i) = \frac{\frac{s_i}{\gamma_i}}{\sum_{j \in I} \frac{s_j}{\gamma_j}} \quad (4)$$

### 2.4.1 Impact of Rate Diversity

To understand the impact of rate diversity, let's assume that each node uses the same packet size, i.e. $\forall i, j \in I, s_i = s_j$. Therefore, based on Equations 2 and 3,

$$T(i) = \frac{\frac{1}{\gamma_i}}{\sum_{j \in I} \frac{1}{\gamma_j}} \quad (5)$$

$$R(i) = \frac{1}{\sum_{j \in I} \frac{1}{\gamma_j}} \quad (6)$$

$$R(I) = \frac{n}{\sum_{j \in I} \frac{1}{\gamma_j}} \quad (7)$$

Equation 6 clearly shows that the throughput of each node $i$ is the same. Thus, under these conditions, DCF achieves throughput-based fairness. Observe, however, the amount of throughput is dependent on the baseline throughputs of all nodes in $I$, which in turn depend on their data rates and packet sizes.

The channel occupancy time $T(i)$ of node $i$ is inversely proportional to the baseline throughput of node $i$, which increases with the increase in transmission rate. Thus, as expected, nodes with slower data rates occupy the channel much longer than those with higher data rates, leading to degradation in the overall network performance.

### 2.4.2 Impact of Packet Size Diversity

The impact of packet size diversity can be understood by assuming that each node uses the same data rate, i.e $\forall i, j \in I$, $d_i = d_j$. Based on Equations 2 and 3, we have:

$$T(i) = \frac{\frac{s_i}{\gamma_i}}{\sum_{j \in I} \frac{s_j}{\gamma_j}} \quad (8)$$

$$R(i) = \frac{s_i}{\sum_{j \in I} \frac{s_j}{\gamma_j}} \quad (9)$$

$$R(I) = \frac{\sum_{i \in I} s_i}{\sum_{j \in I} \frac{s_j}{\gamma_j}} \quad (10)$$

Once again, $R(i)$ depends on the baseline throughputs of all other competing nodes. However, the equations make it clear that in this case $T(i)$ and $R(i)$ may differ across nodes, depending upon packet size.

## 2.5 Impact of the AP Queuing Scheme

The queuing mechanism at the AP dictates the channel bandwidth allocation to clients in the downlink direction. Since the channel conditions at the clients vary, different data transmission rates are often used for different clients. As far as we know, the existing literature on scheduling schemes [20, 22, 27] does not consider the impact of rate diversity. Thus, the aggregate network throughput when only downlink traffic is present is impacted in the same way as previously explained. We also note that if loss rates experienced by nodes differ and both packet transmissions in uplink and downlink directions use different data rates, the achieved throughputs of competing clients may not be equal or easily predictable, even when all nodes use DCF and the AP employs a fair queuing scheme.

## 2.6 Impact of Time-based Fairness

Under the time-based fairness, our proposed definition of fairness, each node achieves an equal share of channel occupancy time. Thus,

$$T'(i) = \frac{1}{n} \quad (11)$$

Substituting Equation 11 in Equations 2 and 3,

$$R'(i) = \frac{\gamma_i}{n} \quad (12)$$

| $d$ (Mbps) | $s$ (Byte) | $n = |I|$ | $\gamma(d,s,I)$ |
|---|---|---|---|
| 11 | 1500 | 2 | 5.189 |
| 5.5 | 1500 | 2 | 3.327 |
| 2 | 1500 | 2 | 1.493 |
| 1 | 1500 | 2 | 0.806 |

Table 2: The experimentally achieved total throughput (or the baseline throughput) of the two nodes simultaneously exchanging data at the same data rate $d$ and packet size $s$. Each node has a similar frame loss rate of less than 2%.

| Fairness Criteria | $R(n1)$ (1) | $R(n2)$ (2) | $R(n3)$ (11) | $R(n4)$ (11) | Total |
|---|---|---|---|---|---|
| RF | 0.436 | 0.436 | 0.436 | 0.436 | 1.742 |
| TF | 0.202 | 0.373 | 1.30 | 1.30 | 3.175 |

Table 3: Comparison of achieved throughputs (in Mbps) of four nodes, each transmitting at 1, 2, 11 and 11 Mbps respectively, under RF and TF. Note that $R(n1)$ under TF is the same as what $n1$ would achieve if all $n2$, $n3$ and $n4$ transmit at 1 Mbps.

$$R'(I) = \frac{1}{n} \sum_{i \in I} \gamma_i \qquad (13)$$

Notice that $R'(i)$ only depends on what node $i$ can achieve under the given conditions and the number of competing nodes. It does not depend upon the data rates or packet sizes used by competing nodes. Unlike $R(I)$ shown in previous subsections, $R'(I)$ is a simple summation of each node's maximum achievable throughput when all competing nodes use its data rate and packet size. $R'(I)$, and $R(I)$ in Equations 7 and 10 will be equal if and only if all nodes in $I$ use the same data rate and packet size.

### 2.7 Examples

In this section we illustrate the ramifications of the differences between Equation 12 and Equation 6 with a small example.

Table 2 shows the experimentally derived baseline throughputs of two identical competing nodes as a function of transmission rate. This provides an estimate of baseline throughput for various transmission rates.

Using these values, we compute the throughputs when $I$ contains four competing nodes, one communicating at 1 Mbps, one at 2 Mbps, and at 11 Mbps. These are shown in Table 3. The achieved throughput of the slower nodes is less under time-based fairness than under throughput-based fairness. Under time-based fairness, the 1 Mbps and 2 Mbps nodes achieve the throughput they would have achieved of all four nodes were running at their speed. The 11 Mbps nodes achieve considerably higher throughput under time-based fairness, and the total throughput improves by 82%.

## 3 Existence of Rate Diversity

In this section, we discuss in detail i) whether rate diversity exists in today's 802.11b networks and ii) whether a single user or multiple users are actively exchanging data during the intervals in which the network is saturated.

To investigate the prevalence of rate diversity, we collected traces of wireless network traffic at one-day Iris student workshop at MIT. There were about 45 attendees and more than half turned on their wireless laptops. We set up a laptop to sniff data during each of the three 90-minute sessions, WS-1, WS-2 and WS-3, all of which took place in a single room of about $40' \times 25'$.

Figure 1 shows the fractions of data bytes transferred using each of the four possible rates during each session. It is clear that rate diversity exists even in a relatively small room. During WS-2, more than 30% of the data bytes were transferred using data rates lower than 11 Mbps.

We also set up an experiment to investigate how an AP change data rates to various clients in indoor office environments. We placed a Cabletron Roamabout-2000 AP in a $18' \times 14'$ office $7'$ above ground. A sender with a wired connection to the AP sent unicast UDP data packets at the saturation rate simultaneously to four different receivers. The first node was about $4'$ away from the AP, the second $12'$ and one thin, wooden wall away, the third $26'$ and two thin wooden walls away and the fourth $30'$ and two thick walls in between. As shown in Figure 1 (see EXP-1), more than 50% of the bytes were transferred using the lowest data rate.

In fact, a recent extensive wireless network usage study on a university campus has found that the average received signal strength varies widely even among positions that are within $20'$ of an access point [19]. Thus, we believe that rate diversity is prevalent in many indoor WLANs and its impact would be much more pronounced with mixed deployments of 802.11b and 802.11g networks.

The negative impact of rate diversity is significant only if the following two conditions are true: i) more than one competing node exchange data during the periods in which network is saturated and ii) competing nodes use diverse data rates. Our analysis of this particular workshop trace data, however, shows that the network is well over-provisioned with 7 APs providing a combined channel capacity of 33 Mbps. However, recent studies have shown that in many enterprise networks [2] and university residential halls [18], WLANs carry significant traffic and contain many APs that have a lot of busy or congested periods.

We analyzed wireless tcpdump trace of Whittemore, a residential facility in the Dartmouth business school where students were required to own laptops. This data was collected by Kotz et al. over the Spring semester [18] and was made publicly available by Kotz. Unfortunately, the trace data does not contain the data transmission rate used for each frame transmis-sion. Nonetheless, we can identify the busy periods in which an AP is carrying close-to the maximum amount of data, and investigate whether more than one user actively exchange data during congested periods.

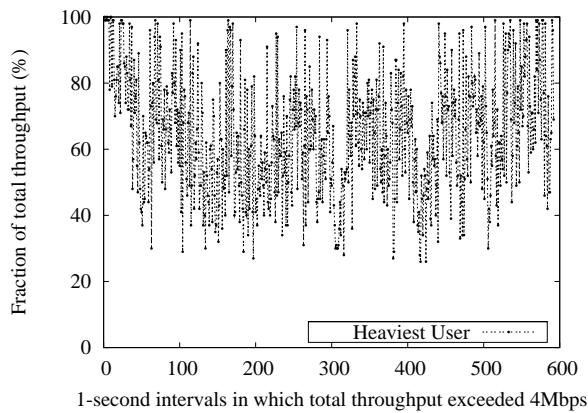

Figure 5: The fraction of throughput achieved by the heaviest user at a busy AP during busy 1-second intervals.

Since TCP dominated the traffic, we conservatively define *busy or congested intervals* as those in which the total data throughput at the AP exceeded 4 Mbps, 80% of the commonly observed TCP saturation throughput when nodes transmit at the maximum data rate and experience a very low loss rate of 1% to 2%.

Figure 5 plots the fraction of aggregate throughput achieved during busy 1-second intervals by the heaviest user at an AP at Whittemore on 8 April 2002, a Spring Monday. The heaviest user is one that exchanged the most bytes with the AP. Although the majority of bytes were transferred by one user on average, it is clear that the heaviest user alone rarely saturated the channel. In most 1-second busy intervals, users other than the heaviest user exchanged significant amounts of data.

## 4 Time-based Regulator

In the previous sections, we have argued that competing nodes should be given an equal amount of long-term channel occupancy time. As explained before, in AP-based WLANs, the MAC protocol and the queuing scheme at the AP in combination determine the channel time allocation. Therefore, to achieve a desired channel time allocation, coordination is necessary between the MAC protocol and the queuing scheme. Our proposed Time-based Regulator runs at each AP, coordinates with clients when necessary and works in conjunction with any MAC protocol.

TBR provides an equal share of long-term channel occupancy time to each competing client node by

- Dictating how packet transmissions are scheduled at the AP as well as at the clients

```
PROCEDURE ASSOCIATEEVENT(i) {
    tokens_i ← T^{init}
    bucket_i ← T^{init}
    rate_i ← fair share of channel occupancy time
    initialize queue_i
}
PROCEDURE FILLEVENT(t) {
    for each bucket_i
        tokens_i ← tokens_i + (t * rate_i)
        if (tokens_i > bucket_i)
            tokens_i ← bucket_i
}
PROCEDURE APPTXEVENT(p) {
    i ← destination of p
    enqueue p to queue_i
}
PROCEDURE MACTXEVENT() {
    for each station i starting with nexti
        if queue_i is not empty and tokens_i > 0
            dequeue a packet p from queue_i
            ask the MAC to transmit p
            nexti ← next station after i
}
PROCEDURE COMPLETEEVENT(p) {
    t ← channel occupancy time of p
    if p was sent by AP
        i ← destination of p
    else
        i ← source of p
    tokens_i ← tokens_i − t
    if (actual_i = 0)
        start_i ← current time
    actual_i ← actual_i + t
}
```

Figure 6: Pseudo-code of TBR

- Taking into account the channel occupancy time of traffic in both downlink and uplink directions, and

- Taking into account varying traffic conditions, loss rates, data rates, and frame sizes

A typical implementation of TBR requires no modification to the underlying MAC protocol and to the drivers of mobile clients, allowing incremental deployment and preserving backward compatibility. Modifications to the clients, however, are necessary to preserve correctness in cases where the uplink UDP flows make up a significant fraction of the WLAN traffic. We will discuss more on this issue in the next subsection.

TBR is based on the leaky bucket scheme [3]. The fundamental unit or token used in the implementation is the channel occupancy time in terms of micro-seconds. TBR only schedules the transmission of a packet destined to or originated from a client only if the node has not used up all its available channel time.

Figure 6 shows the pseudo-code of TBR that runs on the AP. TBR sits above the MAC layer and below the network layer and is implemented in five event handlers, each of which is triggered by the upper layer, timer or the MAC layer.

When a node $i$ associates with the AP (i.e. joins the network), ASSOCIATEEVENT is triggered. The procedure i) creates output queue $queue_i$ and ii) initializes $tokens_i$, the available *tokens*, $bucket_i$, the maximum amount of tokens that the node can accumulate, and $rate_i$, the rate at which tokens are being re-filled.

Whenever the upper layer has a packet $p$ to transmit, it calls APPTXEVENT. TBR simply enqueues the packet to $queue_i$ where $i$ is the destination of $p$.

TBR adjusts $tokens_i$ according to the channel occupancy time of transmitted frames originated from or destined to node $i$. Section 4.2 described how TBR computes the channel occupancy time. $bucket_i$ determines the maximum length of the burst period in which node $i$ can transmit successively (if no other nodes can transmit). $bucket_i$ can affect the short-term fairness and we discuss this issue later in Section 4.5.

TBR sets up a timer that periodically calls FILLEVENT, which for each node $i$, updates $tokens_i$ according to $rate_i$ and $t$, the time elapsed since the last time FILLEVENT was called. $rate_i$ is the rate at which tokens are being re-filled. We note that $\sum_{i=1}^{n} rate_i = 1$, where $n$ is the number of active client nodes. In general, $rate_i$ can vary among client nodes depending on the desired fairness policy. If each competing node should receive an equal share of the channel occupancy time, $rate_i = \frac{1}{n}$. However, in practice, not all nodes can consume their available channel time according to the allocation. TBR ensures that the system remain work conserving by adjusting the token rates appropriately as discussed in Section 4.3.

## 4.1 Scheduling Frame Transmissions

Whenever the MAC layer is ready to accept a new packet for transmission, it calls HWTXEVENT. TBR decides which backlogged packet to release as follows. TBR chooses one output queue among all the output queues with positive available channel time (tokens) and dequeues a packet for transmission.

The manner in which the output queue is chosen has no impact on the overall correctness since only the queues with positive tokens are considered. Nonetheless, the order could impact the short-term fairness. For simplicity and to alleviate short-term unfairness, TBR chooses the output queue among those with positive tokens in a round-robin manner. We note that short-term unfairness can further be reduced by choosing the queue which has the packet with the shortest potential final completion time as in traditional fair queuing schemes [8, 24].

Once the output queue is chosen, TBR can decide which frame in the queue gets transmitted. For TCP, in-order packet delivery is desirable and thus first-in-first-out discipline is preferable. However, if there are time-sensitive packets (used by real-time protocols), they should have priority over TCP packets with earlier arrival times. The correctness of TBR does not depend on how a packet to dequeue is chosen. We also note that TBR works with any buffering scheme (e.g. RED, droptail), whose goal is to decides which packets to drop when the queue is getting full. Note that we distinguish buffering schemes from packet scheduling schemes. The former is responsible for deciding which packets to drop whereas the latter decides which packet gets transmitted [8].

TBR also dictates the scheduling of packet transmissions at the clients. Specifically, whenever $tokens_i \leq 0$, TBR needs to explicitly inform node $i$ to delay transmission for a short amount of time. This can be accomplished in two major ways. First, the TBR agent at the AP informs the client by either sending an explicit notification packet or piggyback such information in a downlink packet when possible. Second, the client monitors the total channel occupancy time of packets transmitted and received and transmits only if there is available channel time allocated for the node. To do so, the client only needs to know $rate_i$. However, as we explain in Section 4.3, TBR at the AP may update $rate_i$ depending on the overall traffic conditions and when that happens, TBR needs to inform the client. In both cases, a client agent is necessary at each client to communicate with TBR at the AP. We choose the first method for simplicity.

The actual amount of communication overhead depends on the MAC protocol used. TBR requires a single bit in the MAC header of a data frame transmission to inform the client to delay its transmission for a pre-determined amount of time. In cases where there is only uplink traffic, TBR can still use the same procedure if the underlying MAC protocol (e.g. DCF) employs a *stop-and-go* retransmission strategy. A stop-and-go protocol requires the node receiving a data frame to reply with a synchronous acknowledgment, which can carry the TBR notification bit. Furthermore, if the underlying MAC protocol employs a polling mechanism (such as 802.11's PCF), no explicit communication is necessary since TBR can dictate which node gets polled.

Cooperation from each client is only necessary if the client has uplink UDP flows that represent a significant fraction of its traffic. Studies of WLAN traffic at university campuses [18, 25] and at a multi-day conference [1] show that TCP accounted for more than 90% of bytes exchanged over the WLANs. TCP data packets are paced by TCP *ack* packets ("ack clocking" [13]) sent out by the receiver. In a typical scenario, all TCP data and *ack* packets go through the same AP. Therefore, delaying TCP data (*ack*) packets at the AP has the effect of slowing down the sending rates of downlink (uplink) TCP flows.

We note that our current TBR implementation does not contain the client-side implementation of TBR. As we demonstrate in Section 5, TBR without the client cooperation can effectively provide long-term channel time guarantees for TCP flows in both directions as well as downlink UDP flows.

## 4.2 Computing Channel Occupancy Time

Whenever the MAC layer has either finished sending or received packet $p$, it triggers COMPLETEEVENT. This procedure subtracts the channel occupancy time of $p$ from the tokens associated with node $i$ that is the source or destination of $p$. It also modifies $actual_i$, the actual tokens used since $start_i$. We will explain how TBR uses $actual_i$ in the next subsection.

We now describe how to compute the channel occupancy time for packet $p$. We define *packet transfer time* as the total time required to transfer a data packet at the 802.11 MAC layer, which is typically the sum of i) the transmission time of the data packet, ii) the transmission time of a synchronous MAC-layer *ack* when necessary, iii) propagation delays for both the data and *ack* packets, and iv) the inter-frame idle periods necessary for the sending node to be idle before accessing the channel. Since the MAC-layer may perform retransmissions upon a transmission failure, the channel occupancy time is the sum of the packet transfer time of each transmission until $p$ has successfully been transmitted or dropped as a result of an undeliverable failure. Therefore, failed packets also contribute to the channel occupancy time of the sending node.

Taking into account retransmissions is straight forward in the downlink direction. However, in the uplink direction, the AP is not aware of the exact number of retransmission attempts made by the client stations. Ideally, the underlying MAC protocol should include a retry sequence number field (about 4 bits) in the header to indicate how many retransmissions precede the current packet transmission.

When retransmission information is not available for each packet received and the necessary header modification is not an option, the AP needs to estimate the information necessary to compute the channel occupancy time. We distinguish two types of losses at the AP: one detected at the MAC layer (due to the CRC check failure) and the other at the physical layer. In the former, it is highly likely that the MAC header, whose size is relatively much smaller than the typical payload size, is not corrupted and thus the AP can determine the source address of the failed transmission as well as the transmission rate. We note that the MAC layer header can be made robust against channel errors by transmitting at a lower data rate.

However, if the frame loss is detected at the physical layer, TBR can be aware of the loss but may not know the necessary transmission information. We believe that heuristics can be developed to estimate the transmission information of each loss detected at the physical layer based on i) the number of active clients in the last few dozen milliseconds, ii) the likelihood of each client contending, and iii) their steady state loss rates at the downlink direction. We plan to develop such heuristics in the future.

## 4.3 Keeping Channel Utilization High

When traffic contains a mixture of TCP and UDP flows that have various sending rates (and bottleneck link bandwidth), it is important to correctly determine the amount of channel occupancy time made available to each node. Specifically, TBR needs to adjust $rate_i$ to reflect changing traffic conditions. For instance, the system will be under-utilized if we give each node $\frac{1}{n}$ of the available channel time but some nodes cannot consume all of their available time shares whereas others can consume more if allowed.

TBR periodically adjusts $rate_i$ associated with each node $i$ so that the channel utilization is kept at maximum without violating the max-min fairness constraint [6, 14]. That is, the smallest $rate_i$ in the network must be as large as possible. Subject to this constraint, the second smallest token rate must also be as large as possible.

We note that DCF in conjunction with a simple round-robin queuing scheme at the AP generally achieves the max-min notion of fairness when only TCP flows are involved. Assume that there are 3 uplink TCP flows and that one flow can only consume $\frac{1}{5}$ of the channel bandwidth (the wireless hop is not its bottleneck link). DCF will allow each of the remaining flows to consume $\frac{2}{5}$ of channel bandwidth provided that the bottleneck link of both flows is the wireless link.

TBR with any MAC protocol achieves the same fairness criteria provided that the MAC layer has the work conserving property that DCF does, i.e. each client node with data to transmit contends for channel access opportunistically. Notice that the max-min fairness criteria does not require that the actual demand of each node is known. Rather, one can simply achieve the fairness goal by incrementally giving more channel time to each competing node that can consume all the channel time made available to it [3]. We implement this general idea in TBR.

Initially each competing node starts with the desired token rate of $\frac{1}{n}$. TBR schedules a timer event called ADJUSTRATEEVENT that periodically adjusts the token *rate* available to each node. As shown in Figure 7, ADJUSTRATEEVENT computes the excess capacity of the *under-utilized nodes*, each with the actual token rate ($actual_i$) lower than the assigned rate by the threshold $R^{th}$. It then computes the excess capacity $E^{min}$ to redistribute equally among nodes ($I'$) that have fully utilized the provisioned bandwidth in the previous round.

The actual method of computing $E^{min}$ is of little importance for the long-term correctness so long as $E^{min}$ is not too big. However, $E^{min}$ does affect the responsiveness of TBR to changing traffic conditions. We will discuss more about this in Section 4.5. If $E^{min}$ is too large, the instantaneous throughputs experienced by flows can significantly vary. Such behaviors may increase the buffer requirements at the nodes to avoid TCP *ack* compression that can lead to packet drops.

Figure 7 shows a particular way of choosing $E^{min}$. We pick,

```
PROCEDURE ADJUSTRATEEVENT() {
    for each node i
        excess ← rate_i - actual_i/(now - start_i)
        if (excess ≤ R^th)
            if excess < E^min)
                E^min ← d
            if excess > E^max)
                m ← i
        else
            add i to set I'
    E^min ← E^min ÷ 2
    for each node j ∈ I'
        rate_j ← rate_j + E^min/|I'|
    rate_m ← rate_m - E^min
    for each node j ∈ I
        actual_j ← 0
}
```

Figure 7: Pseudo-code of the token rate adjustment event

among all under-utilized nodes, node $m$ with the maximal excess capacity (the largest difference in actual and assigned token rate). Half of $E^{min}$ is subtracted from $m$'s token rate and the other half redistributed among nodes that have consumed tokens at rates close to their assigned rates. In Section 5, we show that TBR is able to keep the channel utilization high in the presence of varying traffic conditions.

### 4.4 An 802.11-based Implementation

We implemented TBR in the HostAP [15] driver running on a Linux PC as a proof of concept. The HostAP driver implements access point functionality so that PCs equipped with popular Prism chipset based 802.11 cards can act as APs. We use unique 6-byte MAC addresses as node identifiers.

We note that TBR requires APs to set up per-node output queue. However, the total buffer space requirement is comparable between a normal AP and an AP with TBR. For instance, if an existing AP has the total queue size of $x$ packets than a TBR-equipped AP can setup $n$ queues each with $\frac{x}{n}$ packets, where $n$ is the number of competing nodes. For ease of implementation, our TBR implementation uses FIFO queues. As explained before, TBR can work with any buffering scheme.

Finally, we note that the current implementation of TBR does not use the retransmission information in computing the packet transfer time but we plan to do so in the future. Thus, TBR in some cases can cause slight biases in granting channel occupancy time to competing nodes. Nonetheless, as we show in Section 5, it does well in achieving its goal.

### 4.5 Discussion

TBR is currently intended for ensuring that each competing node receives an equal share of channel occupancy time based on max-min fairness over the long run. As we later demonstrate in Section 5, TBR works well when competing flows last for hundreds of packets.

Although we believe that long-lived flows (e.g. file transfer applications) are usually the cause of congestion in enterprise and university networks, we acknowledge that congestion in *hotspot* access networks may be caused by many short-lived flows with diverse data rates, each sending only dozens of packets.

Responsiveness of TBR relies on how it adjusts the token rate assigned to each competing node and how often (see ADJUSTRATEEVENT). Furthermore, the burst period ($bucket_i$) in which node $i$ can transmit successively also influences the responsiveness of TBR as well as short-term fairness. Special attention must be paid to a packet-level interaction between TBR and the underlying MAC so that TBR can respond to varying traffic conditions in the order of tens of packet transfer time. In the future, we plan to understand each of these issues in detail and make TBR responsive for very short-lived flows as well.

Large $bucket_i$ can exacerbate the short-term unfairness, i.e. some competing nodes do not achieve their desired fair shares within a very short interval, commonly found in 802.11 WLANs [17]. Short-term unfairness in its most severe form leads to TCP ack compression in which multiple TCP acks arrive at the sender, which then sends several TCP packets successively, leading to undesirable packet drops at the bottleneck queue. However, the TCP ack compression problem can be effectively solved by pacing TCP packets [5].

TBR can potentially be modified to provide each competing node with the desired share of channel occupancy time (not necessarily equal). Therefore, QoS mechanisms may use TBR to provide QoS at existing AP-based WLANs. We also note that although the current implementation of TBR allocates channel time to nodes, it can be extended to allocate channel time among various flows of each node.

We note that the 802.11e standard [12] currently being drafted defines quality of service support for the 802.11 MAC. Using 802.11e, competing nodes acquire Transmission Opportunities (*TXOP*), each of which is defined as an interval of time when a station has the right to initiate transmissions. TXOPs are allocated via contention or granted through the centralized coordinator like the AP. 802.11e differentiates the probability of channel access based on the traffic categories. TBR can be integrated with 802.11e by choosing appropriate traffic categories for each competing node according to their fair share of channel occupancy time.

## 5 Evaluation

We setup experiments to evaluate the correctness and performance of TBR. We used a PIII-700MHz Linux laptop equipped with a D-Link DWL-650 card running the Hostap driver as the AP and IPAQs equipped with Cisco-350 cards as competing nodes.

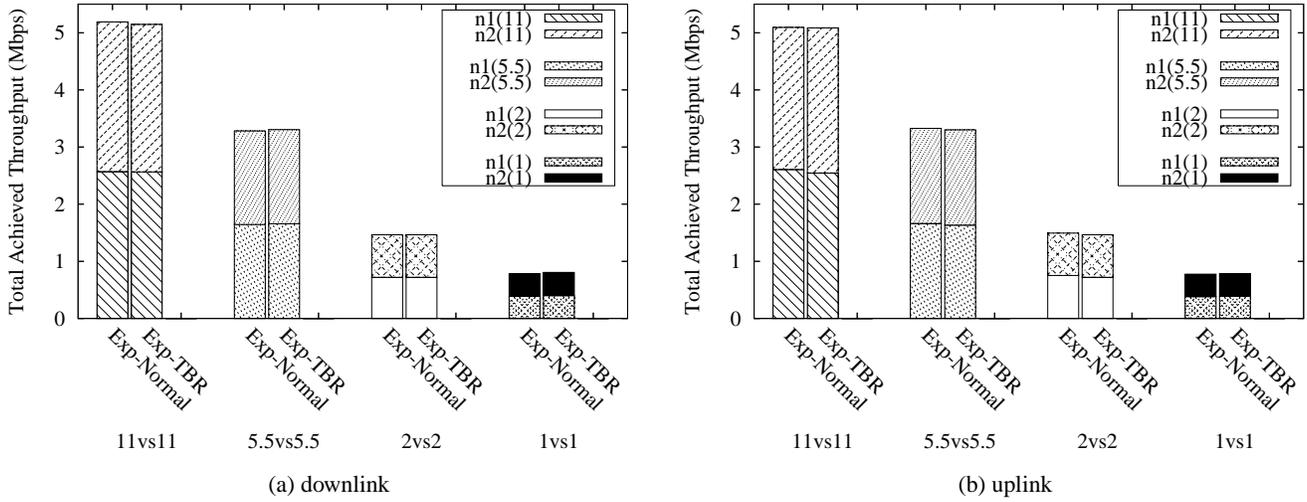

Figure 8: TCP throughputs achieved in either uplink or downlink direction by two competing nodes using the same data rate. *Exp-Normal* and *Exp-TBR* denote the experiments that were run with the AP equipped without or with TBR respectively. *n1(11)* denotes the throughput achieved by node *n1* transmitting at 11 Mbps.

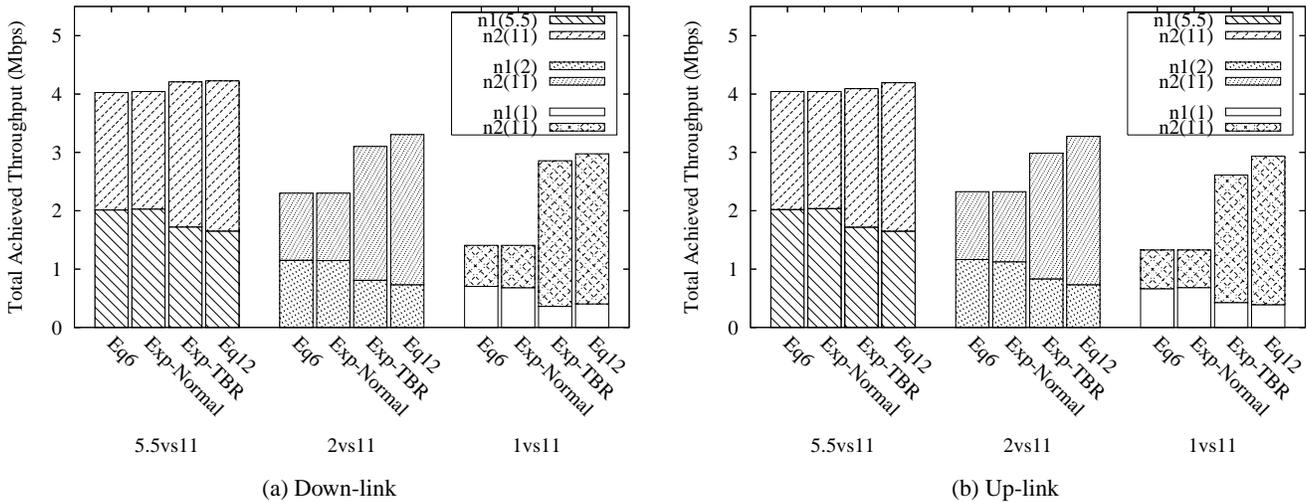

Figure 9: TCP throughputs achieved in either up-link or down-link direction by two competing nodes using different data rates. *Exp-Normal* and *Exp-TBR* denote the experiments that were run with the AP equipped without or with TBR respectively. *Eq6* and *Eq12* represent the achieved throughputs according to Equation 6 and Equation 12 respectively. *n1(11)* denotes the throughput achieved by node *n1* transmitting at 11Mbps.

For each type of experiment, we ran in two different AP configurations: one with TBR, *Exp-TBR*, and one without, *Exp-Normal*. Each data point is an average of 5 to 10 runs and in each run, each contending node sends about 2000 1500-byte packets. All throughputs measured are achieved TCP throughputs.

When the AP is run under the normal configuration, no queue is set up in the driver. Instead, the kernel interface queue (with the maximum size of 110) is used to store packets. When the AP is run with TBR, $n$ queues each with the maximum queue size of $\frac{100}{n}$ is set up inside the driver. The kernel interface queue is then set to 10. Thus, the total buffer space available to each scheme is the same.

Figure 8 compares the throughputs achieved by two competing nodes when the AP is configured with or without TBR. When competing nodes use the same data rate, *Exp-TBR* and *Exp-Normal* yield almost identical results, showing that TBR incurs little overhead.

When nodes use different data rates, the throughput achieved by each competing node as well as the total throughput differ significantly depending upon whether TBR is used or not. As shown in Figure 9(a), when TBR is used, the total achieved throughput in the down-link direction increases by about $6\%$ in the 5.5vs11 case, $35\%$ in the 2vs11 case and $103\%$ in the 1vs11 case.

Analytical (*Eq6*) and experimental (*Exp-Normal*) values agree for all the cases when the AP is configured without TBR. Similarly, *Exp-TBR* and *Eq12* show very similar results, affirming that our regulator achieves the objective of providing long-term equal channel occupancy time to competing nodes. The slight differences in performance between *Exp-TBR* and *Eq12* is due to the fact that TBR needs to estimate channel occupancy time without the retransmission information available. Whenever a packet loss is experienced by a node, the channel occupancy time of that node needs to be decreased accordingly. Without the retransmission information, TBR in this case slightly biased the node sending at a lower data rate, thus decreasing the total throughput by a small amount compared to *Eq12*. In the future, we plan to extract (from the card firmware) or estimate retransmission information as suggested in Section 4.

Figure 9(b) shows similar improvements achieved by TBR in the up-link direction. We also ran experiments involving mixed up-link and down-link TCP flows and found similar results (not shown here).

| Throughput | Exp-Normal | Exp-TBR |
|---|---|---|
| n1 | 2.9434 | 2.9542 |
| n2 | 2.1276 | 2.1193 |
| Total | 5.071 | 5.061 |

Table 4: Comparison of achieved TCP throughputs under Exp-Normal and Exp-TBR. Node $n2$ experienced the bottleneck bandwidth of 2.1 Mbps whereas node $n1$ could send as fast as it could (TCP permitted). Both nodes transmitted at 11 Mbps.

To understand how well TBR works when traffic contains flows with various demands, we set up a scenario that involved two nodes, *n1* and *n2*, each sending TCP packets at the same data rate of 11 Mbps but experienced different bottleneck link capacities. *n2* experienced the bottleneck bandwidth of 2.1 Mbps while the wireless link is *n2*'s bottleneck. We achieved this by limiting the sending rate of the application generating TCP packets at *n2*. The expected DCF's behavior is to give *n2* 2.1 Mbps of channel bandwidth and *n1* the remainder. Table 4 shows the throughputs achieved under *Exp-TBR* and *Exp-Normal*. There is no significant difference between the two sets of results showing that the rate adjustment algorithm described in Section 4.3 works.

## 6 Related Work

We note that the general idea of temporal sharing in the context of multi-rate WLANs has been mentioned before by Sadeghi *et al.* [23]. They have proposed an opportunistic rate adaptation scheme (called OAR) that achieves significant throughput gain over previously proposed rate adaptation schemes [11, 16]. The key idea behind OAR is to allow nodes that have high-quality channel condition to transmit more than one packet at a time taking advantage of time-correlated channel conditions. OAR simply allows a node that can transmit at 11 Mbps 5 times more opportunities than the node transmitting at 2 Mbps. OAR justifies this by saying that nodes are achieving similar time-shares as when they both are transmitting at 2 Mbps. OAR is a DCF-based protocol mainly intended for *ad hoc* networks and requires modifications to DCF. Unlike AP-based networks, *ad hoc* networks, in which nodes communicate with each other without using access points, are more suitable when communications among wireless nodes are dominant or no wired infrastructure exists. In contrast, AP-based networks are designed for communications among wireless nodes and other nodes that can be reached via a wired infrastructure to which APs are connected.

Unlike the previous work, we investigate and explain the differing impacts of the fairness notions on the network performance and our work focuses on AP-based 802.11 networks in which the queuing scheme at the AP significantly impacts the channel capacity allocation.

Recently, Heusse *et al.* have shown through simulations and experiments that performance degradation occurs when two nodes are sending at different data rates [10]. Through analysis, authors show that the node sending at a lower data rate will achieve the same throughput as other nodes sending at higher data rate. The authors do not suggest any mechanism to mitigate this effects.

Efforts have been made in developing distributed fair scheduling algorithms that are suitable for the shared wireless medium. [20, 22, 27]. Like the schemes proposed in wired networks [8, 9, 24], these wireless scheduling algorithms [20, 22, 27] neither take into account the impact of transmission rate diversity nor the channel resource for both downlink and uplink traffic as most schemes [22, 27] were targeted for *ad hoc* wireless networks.

## 7 Summary and Conclusion

We started by showing that, in the presence of rate diversity, the throughput-based fairness notion implemented by the 802.11's popular MAC protocol and the traditional queuing schemes at the APs leads to a situation in which the aggregate throughput is determined largely by the slowest node.

We next presented a time-based notion of fairness that provides an equal amount of long-term channel occupancy time to each competing node. This prevents faster nodes from being dragged down by slower ones. Moreover, it satisfies what we called the *baseline property*, i.e., the achieved throughput of any competing node in a multi-rate WLAN is equal to what it would achieve in a single-rate WLAN in which all competing nodes transmit at its data rate. In the presence of rate diversity, using this definition of fairness can lead to vastly improved aggregate network throughput, more than $100\%$ in some realistic scenarios.

We next described a practical scheme called TBR that works in conjunction with any MAC protocol to provide long-term time-based fairness in AP-based WLANs by appropriately scheduling packet transmissions. We showed that TBR can be implemented in an AP driver in a way that is backwards compatible with existing 802.11 standard. We implemented our scheme in the Linux Hostap driver running on a PC used as the AP, and evaluated it through a series of experiments. In the absence of rate diversity, the performance of our implementation is equivalent to the standard implementation. In the presence of rate diversity, it achieves the predicted gains.

In today's AP-based 802.11b WLANs, rate diversity is already common as our trace analyses show. As newer standards such as 802.11g are deployed, the problem will become worse. For an extended period of time 802.11 WLANs will run in a mixed mode, and if 802.11g clients are slowed down to run at the rate of 802.11b clients, there will be little incentive to upgrade. We believe that switching to time-based fairness is a good option.

## References

[1] A. Balachandran, G. M. Voelker, P. Bahl, and P. V. Rangan. Characterizing user behavior and network performance in a public wireless LAN. ACM Press, June 2002.

[2] M. Balazinska and P. Castro. Characterizing mobility and network usage in a corporate wireless local-area network. In *Proc. of ACM MOBISYS'03*, May 2003.

[3] D. Bertsekas and R. Gallager. *Data Networks*. Prentice Hall, second edition, 1992.

[4] J. Bruno, E. G. Coffman, and R. Sethi. Scheduling independent tasks to reduce finishing time. *Communications of the ACM*, 17:382–387, Jul 1974.

[5] M. C. Chan and R. Ramjee. TCP/IP performance over 3g wireless links with rate and delay variation. In *Proc. of ACM MOBICOM'02*, pages 71–82, 2002.

[6] D.-M. Chiu and R. Jain. Analysis of the Increase/Decrease Algorithms for Congestion Avoidance in Computer Networks. *Computer Networks and ISDN Systems*, 17(1):1–14, June 1989.

[7] Data Sheet of Cisco Aironet 350 Series Access Points. http://www.cisco.com/warp/public/cc/pd/witc/ao350ap/prodlit/carto_in.htm.

[8] A. Demers, S. Keshav, and S. Shenker. Analysis and Simulation of a Fair Queueing Algorithm. *Internetworking: Research And Experience*, 1:3–26, April 1990.

[9] P. Goyal, H. M. Vin, and H. Cheng. Start-time fair queueing: A scheduling algorithm for integrated services packe switching networks. *IEEE/ACM Transactions on Networking*, oct 1997.

[10] M. Heusse, F. Rousseau, G. Berger-Sabbatel, and A. Duda. Performance anomaly of 802.11b. In *Proc. of IEEE INFOCOM'03*, April 2003.

[11] G. Holland, N. H. Vaidya, and P. Bahl. A rate-adaptive MAC protocol for multi-hop wireless networks. In *Proc. of ACM MOBICOM'01*, pages 236–251, 2001.

[12] IEEE 802.11 Working Group. Draft Supplement to International Standard for Information Exchange between systems - LAN/MAN Specific Requirements, Nov. 2001.

[13] V. Jacobson. Congestion avoidance and control. *ACM Computer Communication Review*, 18, 4:314–329, 1988.

[14] R. Jain, D.-M. Chiu, and W. Hawe. A Quantitative Measure of Fairness and Discrimination for Resource Allocation in Shared Computer System. Technical Report 301, Digital Equipment Corporation, Sept. 1984.

[15] Jouni Malinen. Host AP driver for Intersil Prism2/2.5/3. http://hostap.epitest.fi, 2003. Version 0.0.1.

[16] A. Kamerman and L. Monteban. Wavelan ii: A high-performance wireless lan for the unlicensed band. *Bell Labs Technical Journal*, pages 118–133, Summer 1997.

[17] C. E. Koksal, H. I. Kassab, and H. Balakrishnan. An analysis of short-term fairness in wireless media access protocols. In *Proc. of ACM SIGMETRICS'00*, June 2000.

[18] D. Kotz and K. Essien. Analysis of a campus-wide wireless network. In *Proc. of ACM MOBICOM'02*. ACM Press, Sept. 2002.

[19] D. Kotz, C. Newport, and C. Elliott. The mistaken axioms of wireless-network research. Technical Report TR2003-467, Dept. of Computer Science, Dartmouth College, July 2003.

[20] S. Lu, V. Bharghavan, and R. Srikant. Fair scheduling in wireless packet networks. *IEEE/ACM Transactions on Networking*, 7(4):473–489, 1999.

[21] ORiNOCO AS-2000 System Release Note. http://www.michiganwireless.org/tools/Lucent/ORiNOCO/AS-2000_Rel2_1/AS2000_R2_10_01_Readme.txt.

[22] P. Ramanathan and P. Agrawal. Adapting packet fair queueing algorithms to wireless networks. In *Proc. of ACM MOBICOM'98*, pages 1–9, 1998.

[23] B. Sadeghi, V. Kanodia, A. Sabharwal, and E. Knightly. Opportunistic media access for multirate ad hoc networks. In *Proc. of ACM MOBICOM'02*, sept 2002.

[24] M. Shreedhar and G. Varghese. Efficient Fair Queuing using Deficit Round Robin. In *Proc. of ACM SIGCOMM'95*, August 1995.

[25] D. Tang and M. Baker. Analysis of a metropolitan-area wireless network. *Wireless Networks*, 8(2/3):107–120, 2002.

[26] Y. Tay and K. Chua. A capacity analysis for the IEEE 802.11 MAC protocol. *ACM/Baltzer Wireless Networks*, 7(2):159–171, Mar 2001.

[27] N. H. Vaidya, P. Bahl, and S. Gupta. Distributed fair scheduling in a wireless LAN. In *Proc. of ACM MOBICOM'00*, pages 167–178, 2000.